\documentclass[twocolumn,prl,aps]{revtex4}

\usepackage{amssymb}
\usepackage{bm}% bold math
\usepackage{epsfig}
\usepackage{graphicx}
\usepackage{amsmath}
\usepackage{color}

\def\Ef{E$_{\mbox{\tiny F}}$}

 \def\vs{\mbox{V$_s$}}
\def\Vs{\mbox{V$_{s}$}}
\def\It{\mbox{I$_t$}}

\begin{document}

\title{Gating the charge state of a single molecule by local electric fields}
\author{I. Fern\'andez-Torrente$^1$, D. Kreikemeyer-Lorenzo$^{1,2}$, A. Strozecka$^1$,
K.J. Franke$^1$ and J.I. Pascual $^1$}
\affiliation{$^1$Institut f\"ur Experimentalphysik, Freie Universit\"at Berlin, Arnimallee 14, 14195 Berlin, Germany\\
$^2$ Fritz-Haber-Institut der Max-Planck-Gesellschaft, Faradayweg 4-6, 14195
Berlin, Germany\\ }
\date{\today}

\begin{abstract}

The electron acceptor molecule TCNQ is found in either of two distinct integer
charge states when embedded into a monolayer of a charge transfer-complex on a
gold surface. Scanning tunneling spectroscopy measurements identify these
states through the presence/absence of a zero-bias Kondo resonance. Increasing
the (tip-induced) electric field allows us to reversibly induce the
oxidation/reduction  of  TCNQ species from their anionic or neutral ground
state, respectively. We show that the different ground states arise from slight
variations in the underlying surface potential, pictured here as the gate of a
three-terminal device.

\end{abstract}

\maketitle
%\newpage

A goal of molecular electronics is to use single molecules as transistors
\cite{AviramCPL74, ParkNature00, JoachimNature00, SongNature09}. To achieve
this paradigm a high degree of charge localization in the device is required
\cite{NitzanScience03}. Charge localization favors that a molecular device can
only change its electron occupation in integer numbers, leading to discrete
changes of its transconductance \cite{ParkNature02,LinagNature02}. However, a
molecule contacted by metal electrodes generally shows non-integer charge
states, in which the electron density is redistributed throughout the bond with
the leads \cite{NitzanScience03, XuePRB03}. To achieve a large degree of charge
localization, the interaction of the molecule with the metal leads has to be
sufficiently weak. This is best achieved by introducing an "effective tunneling
barrier" to electronically decouple molecular states from the metal. Several
STM studies on single molecules adsorbed on atomically clean and
well-structured metal surfaces have shown that this is achievable using thin
insulating layers (ionic or oxide thin films) separating the molecule from the
surface \cite{WuPRL04, ReppScience04, ReppPRL05, PradhanNanolett05, FuPRL09,
SwartNanoLett11}. A comparable level of decoupling has been observed when
combining two different molecules in a monolayer directly adsorbed on a metal
surface \cite{FrankePRL08, TorrentePRL08}. In this case, the proximity of the
metal surface screens more efficiently charges localized at the molecule,
leading to a reduction of Coulomb charging energy U and, hence, of the energy
cost to change the electron occupation of the molecule \cite{TorrenteJPCM08}.
Here, we find a similar degree of charge localization on the electron acceptor
molecule tetracyanoquinodimethane (TCNQ) mixed with the electron donor
tetramethyl-tetrathiafulvalene (TMTTF) in a stoichiometrically ordered
monolayer. The methylation of the TTF backbone reduces the interaction of the
molecular ad-layer with the metal substrate in comparison to the parent
compound TTF-TCNQ \cite{TorrentePRL08, GonzLakPRL08}, while maintaining ground
states with integer charge occupation.

A second requirement for molecules acting as electronic building blocks is the
tunability of their charge state which would allow us to control the molecular
transconductance  \cite{Kouwenhoven01}. To change the charge state one needs an
external handle (i.e. a potential) that shifts the alignment of molecular
levels around the leads' chemical potential. A successful approach consists in
using the local electric field induced by the proximity of a metal STM tip to
induce such shift \cite{WuPRL2004, PradhanPRL05, NazinPRL05, TeichmannPRL08,
WiesendangerPRB2008}. In this case, the critical field for changing the charge
state can be controlled  by either the applied sample bias or the tip position.

In this Letter, we demonstrate the localization of charge and its manipulation
by an electric field in individual electron-acceptor molecules embedded into a
self-assembled monolayer on a metal surface.  Tunneling transport measurements
reveal the existence of two integer charge states that can be manipulated
reversibly by means of a gating electric field, locally applied using the tip
of a scanning tunneling microscope (STM). Our results suggest that this model
molecular junction can be treated conceptually as a three-terminal transistor.

\begin{figure}[tb] \begin{center}
\includegraphics[width=0.95\columnwidth]{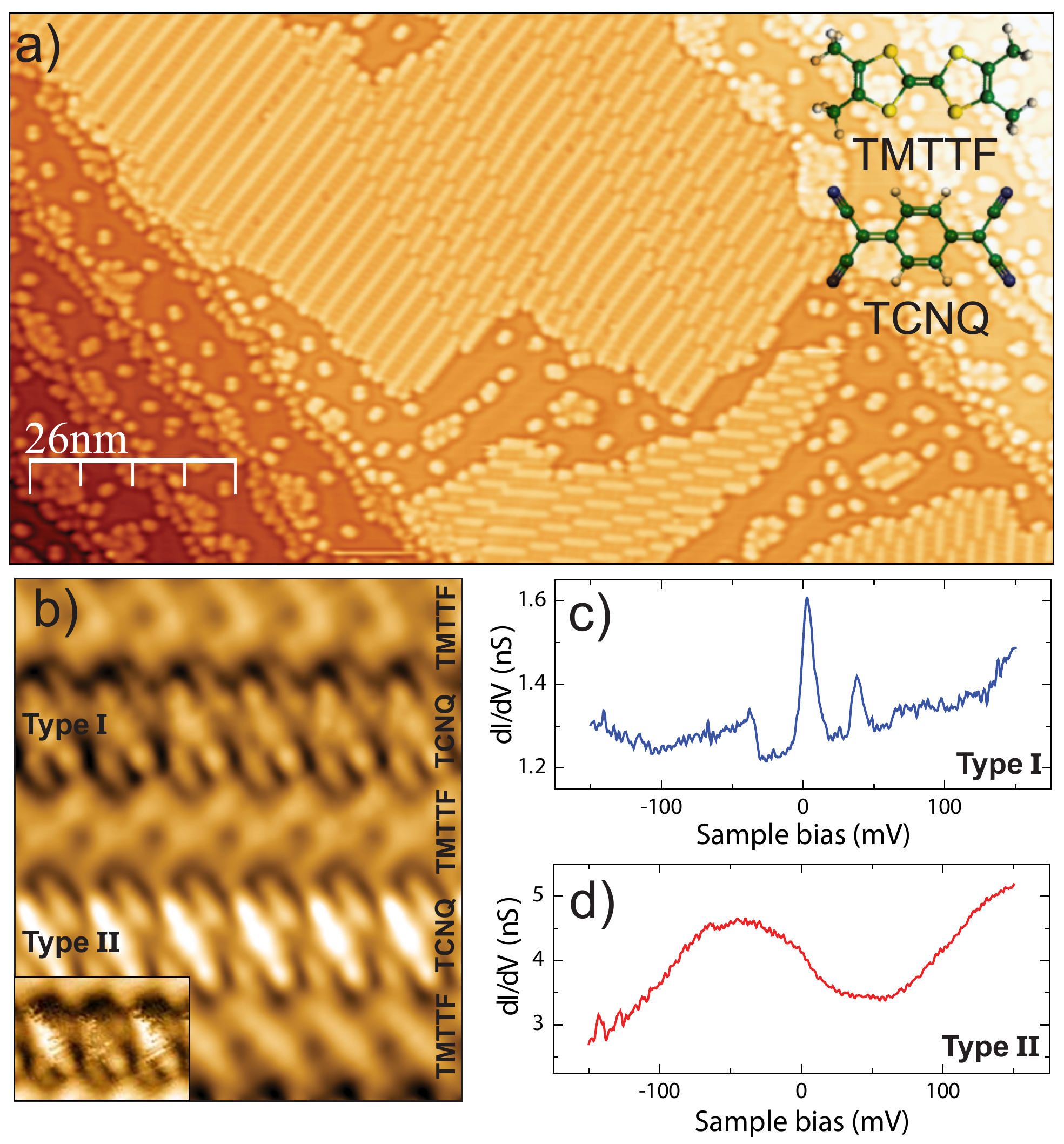}
\end{center}
\caption{(color online) Figure 1: (a) STM image \cite{wsxm} of self-assembled TMTTF-TCNQ domains on Au(111)
(the molecular structure of both species are shown in the image) (\vs = 1 V; \It=0.4 nA). The topographic height difference is $\Delta z \sim$ 20 pm.
(b) Intramolecular structure of TMTTF and TCNQ molecules
resembles the shape of the free molecule HOMO and LUMO, respectively (\vs = 90 mV; \It=0.17 nA). Inset: TCNQ molecules switching under the influence of the scanning tip between Type I and Type II (\vs = 90 mV; \It=0.17 nA).
(c,d) Spectra acquired over the centre of  "darker" (Type I) and "brighter" (Type II) TCNQ
molecules, respectively (\vs = 150 mV; \It=0.5 nA, The lock-in modulation is 2 mV rms at 723 Hz). The peak features in the former correspond to a Kondo
resonance, at \vs=0 mV, with  two sidebands at \vs=$\sim \pm$ 40 mV due to the excitation of a molecular
vibration,  as in Ref. \cite{TorrentePRL08}. The flat spectrum in (d) suggests a neutral ground state. }\label{FigSTM}
\end{figure}

TMTTF-TCNQ monolayers are grown by sequential sublimation of the two compounds
on an atomically clean Au(111) surface kept at room temperature and in
ultra-high vacuum \cite{note1}. STM images, obtained at a base temperature of
4.8 K, show ordered molecular domains composed of electron-donor and -acceptor
species alternating in molecular rows (Fig. 1a and 1b). The molecules interact
among each other through their H (donor)-CN (acceptor) terminal groups, while
keeping their backbone plane parallel to the metal surface. The TMTTF-TCNQ
layer leaves the herringbone reconstruction intact and incorporates numerous
dislocations in its structure (as shown in Fig. 1a). A crucial finding here is
that TCNQ molecules show striking variations of their apparent height in the
STM images at low bias; for example, molecules labeled Type I and Type II in
Fig. 1b show a height difference of $\sim$ 20 pm, and some may even exhibit
bistable behavior between the two heights during STM scanning [inset of Fig.
1b]. We will show next that the differences in apparent height are correlated
with  a different charge state of the TCNQ species.

To identify the electronic ground state of the two types of TCNQ molecules we
use tunneling spectroscopy. Focusing first on the spectral region close to the
Fermi level (\Ef), the differential conductance plots on the darker TCNQ
species (Type I) show a narrow ($\sim$7  mV FWHM) zero-bias peak (ZBP), similar
to that found on the parent TTF-TCNQ compound \cite{TorrentePRL08}. This ZBP is
a manifestation of the spin Kondo effect and reveals that TCNQ molecules have a
S=1/2 magnetic ground state. Such state is created by the occupation of the
TCNQ LUMO with a single electron donated by its environment (here TMTTF and
surface). The Kondo peak measured at the center of the molecule appears split
in two side bands centered at $\pm \sim$ 40 meV, corresponding to the
excitation of an in-plane molecular vibration, which strongly couples with the
TCNQ LUMO \cite{TorrentePRL08}. The spectra of the brighter molecules (Type
II), in contrast, are rather flat without any indication of a Kondo resonance.
They thus lie in a different, presumable neutral, electronic ground state.

\begin{figure}[tb]
\begin{center}
\includegraphics[width=0.9\columnwidth]{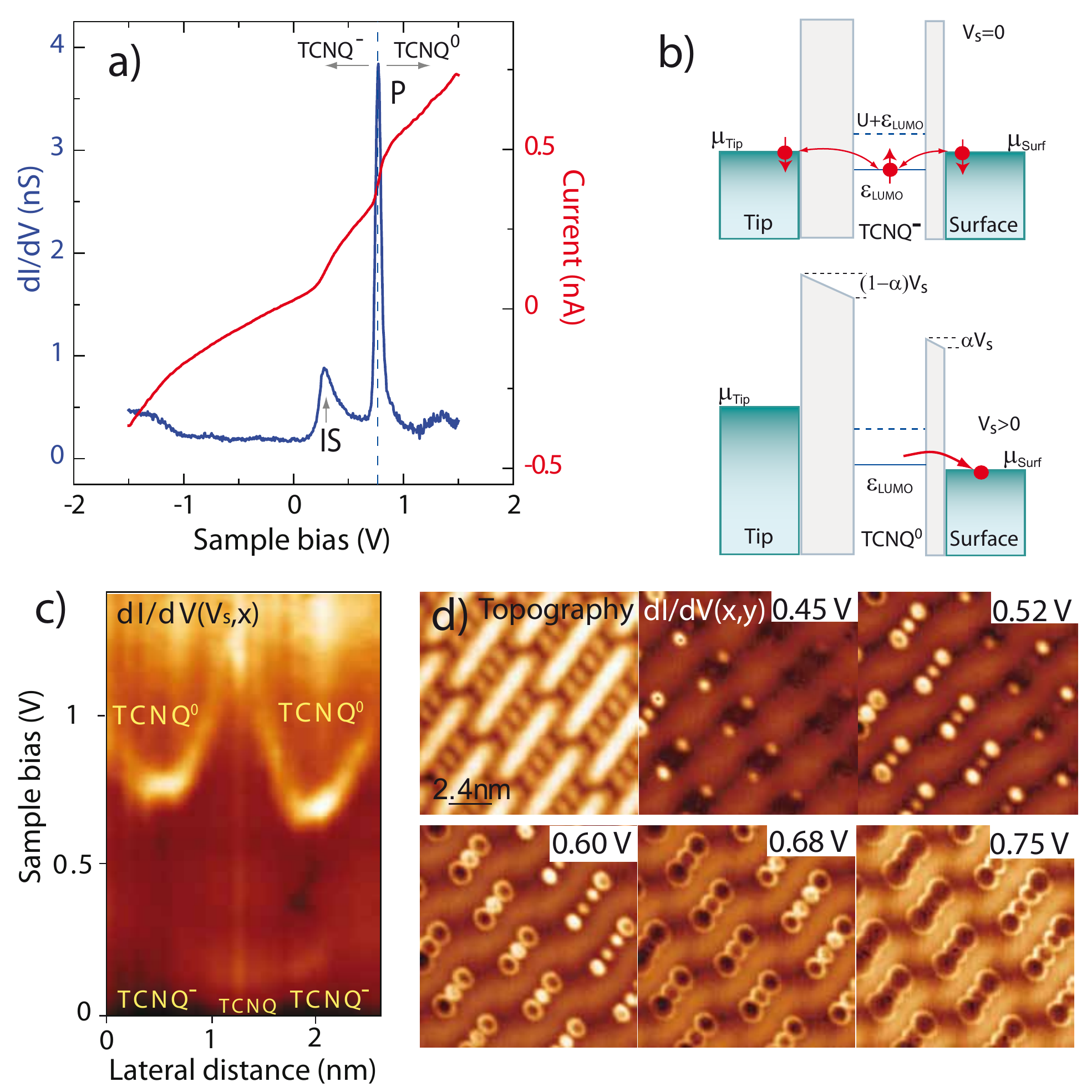}
\end{center}
\caption{(color online) (a) Current and dI/dV spectrum acquired over the centre of Type I molecules. The
spectrum shows the onset of the hybrid metal-organic interface state (IS), and a sharp
peak (P) due to the change of the charge state of the TCNQ molecule, as indicated.
(b) Potential energy model depicting schematically the field-induced
discharging of a TCNQ$^-$ anion at positive sample biass.
(c) Spectral maps of dI/dV vs. \vs and lateral distance, showing the shift of peak P as a
the tip approaches the center of a molecule. (d) Constant current dI/dV maps at different
bias voltages evidence an elliptical equipotential contour  tracing  the onset for
discharging (\It=0.7 nA).} \label{fig2}
\end{figure}

The charge state of the two types of molecules can be manipulated by increasing
the  electric field at the molecular junction.  This is done by ramping the
sample bias (\vs) in a wider bias range. Examples of the simultaneously
recorded current and conductance spectra are shown in Figures 2a and 3a. On
Type I molecules, the conductance plots show two types of features at positive
bias (Fig. \ref{fig2}a): a step-like onset at $\sim$0.2 eV (IS), corresponding
to a (molecular-induced) interface state \cite{note2}, and a sharp "peak" (P).
The position of peak P shifts gradually to lower values as the distance between
tip and molecule is reduced (Fig. 2c), hence, as the electric field at the
molecule is increased. This behavior is qualitatively interpreted by a double
barrier tunneling junction (DBTJ).

In the DBTJ model (Fig. 2b) a small fraction $\alpha$ of the applied bias
potential drops between the molecule and the surface \cite{PradhanPRL05,
NazinPRL05, DattaPRL97}. Correspondingly, the molecular levels shift by an
amount $\alpha$V$_s$ with respect to the surface chemical potential. In Type I
molecules, a positive bias shifts the alignment of the singly occupied TCNQ$^-$
LUMO (i.e. the SOMO) to higher energy. Upon crossing the surface chemical
potential, the SOMO is emptied, leaving the molecule in a neutral state. This
sudden discharging causes a stepped increase of the tunneling current and,
consequently, a sharp peak (P) in the differential conductance \cite{WuPRL2004,
note7}.

The fraction $\alpha$ increases as the tip approaches the molecule, and the
double barrier junction becomes more symmetric. In full agreement with this
model, constant current dI/dV maps in Fig. 2d  show that the "discharging" peak
P is distributed around the TCNQ molecules with an elliptical contour, whose
perimeter increases with the sample bias. These ellipses depict  contours of
constant electric field inducing the discharging  \cite{NazinPRL05,
PradhanPRL05};  the region inside corresponds to the neutral specie,
discharged by the proximity of the STM tip.

\begin{figure}[tb] \begin{center}
\includegraphics[width=0.9\columnwidth]{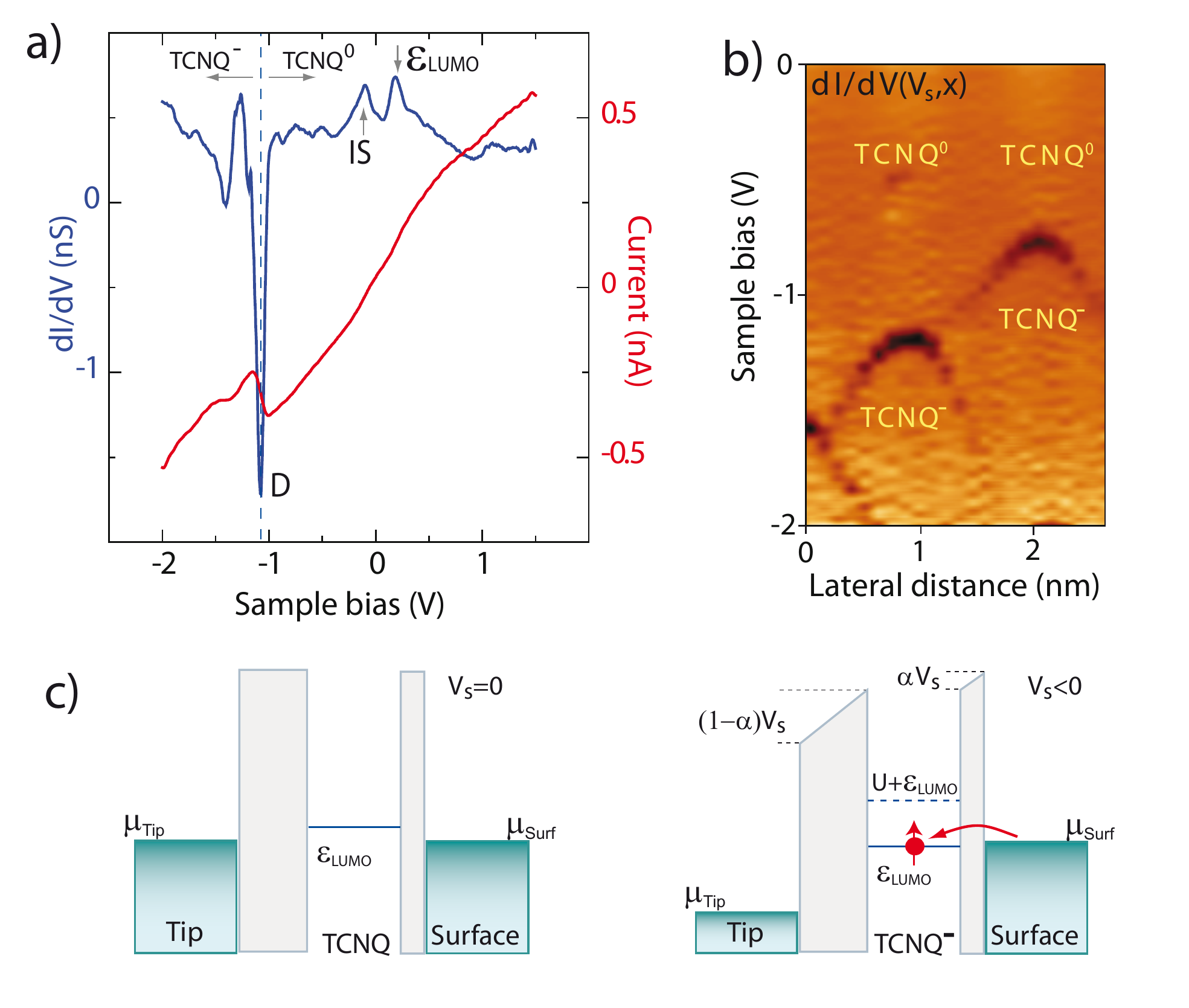}
\end{center}
\caption{(color online) (a) Current and dI/dV spectrum acquired over the centre of Type II molecules. The
interface state (IS) lies below the Fermi level, the LUMO is found above
\Ef. A sharp dip (D) results from the blocking of the tunneling current
upon charging. (b) Spectral dI/dV (\Vs, distance) map showing the shift of
dip D with varying tip-molecule distance. (c) Model of the voltage drop in
the DBTJ explaining the charging of the TCNQ molecules. }
\end{figure}

The spectra of brighter TCNQ molecules in Fig. 1b (Type II) show, in a similar
energy scale, a quite different behavior (Fig. 3a). The most striking feature
is a sharp dip (D) at negative bias, which, as for Type I species, shifts
towards \Ef\ for closer tip distances to the molecule (Fig. 3b). This property
is thus a fingerprint of the opposite effect than peak P: the local charging
(i.e. the reduction) of the TCNQ molecule induced by the field-driven downshift
of their empty LUMO level below \Ef\ (Fig. 3c) \cite{note7}.

The magnitude of the electric field which acts as a gate potential for
molecular discharging/charging is controlled by adjusting the tip position with
respect to the molecule. This control of the charge state thus resembles the
situation in a three terminal device. The change in charge state of the
molecule  leads to a step in the tunneling current following the same trend for
both type I and II molecules: charging a TCNQ molecule causes a partial
blockade of the current (visualized as a step-wise decrease in the current
spectra of Figs. 2a and 3a and the charging onset as the bias is scan from
right to left), while discharging causes a substantial increase in the current.
The lower tunneling transmission through negatively charged molecules is
attributed to the interface dipole created by the anion and its image charge,
resulting in an increase of the local work function \cite{note3}. This is also
consistent with the lower apparent height of type I molecules in STM images at
low bias (e.g. Fig. 1b).

\begin{figure}[tb]
\begin{center}
\includegraphics[width=0.9\columnwidth]{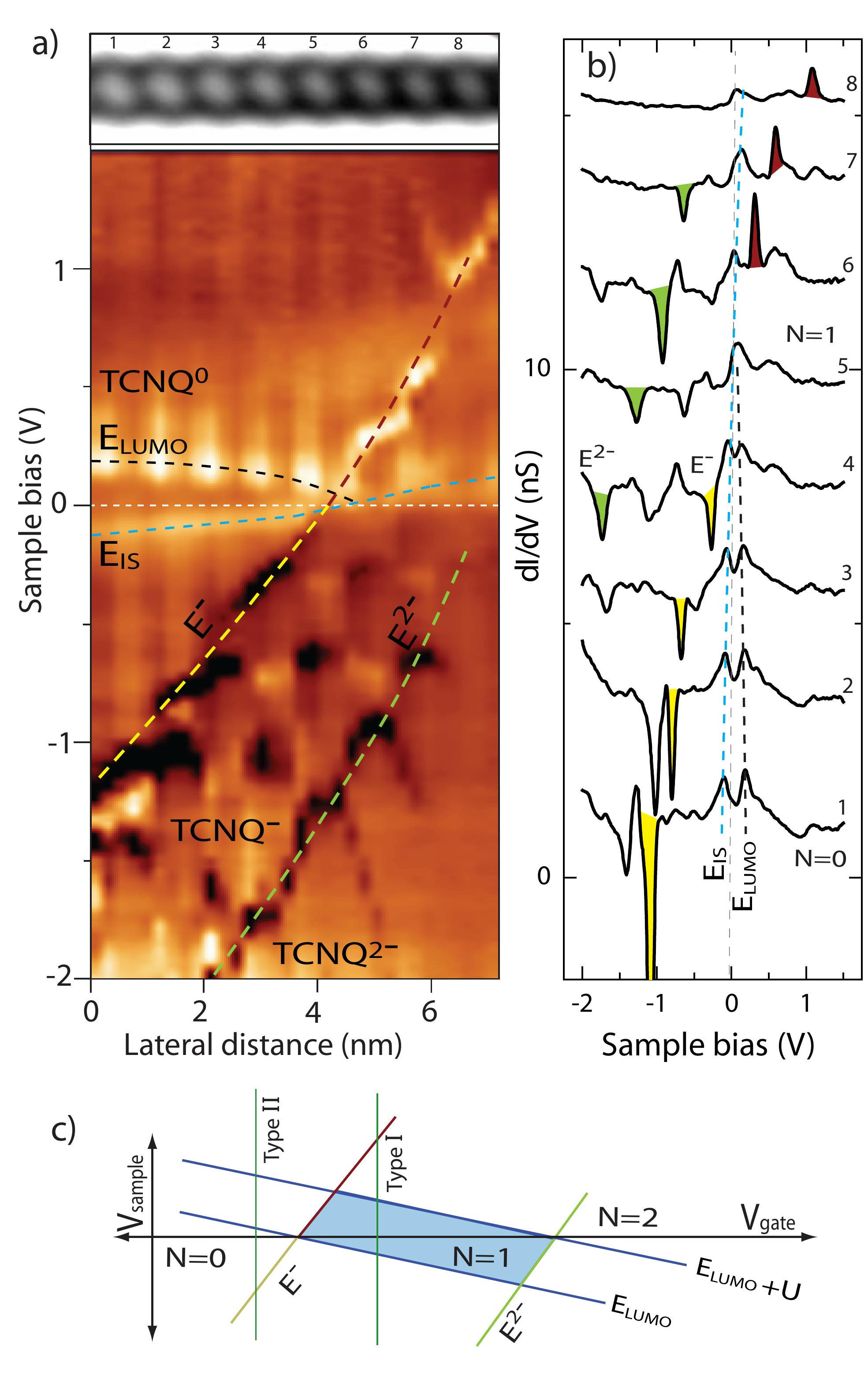}
\end{center}
\caption{(color online) (a) dI/dV vs. distance map along eight TCNQ molecules in an
extended molecular row; (top) STM image and labels the different sites within the  molecular chain (\vs = 1.5 V, \It=0.23 nA).  Dashed lines guide the gradual shift of the charging dip, discharging peak, interface state and LUMO with the molecular site. (b) dI/dV spectra acquired over the centre of the eight molecules.  (c) Scheme of the variation of features in (b) and (c) with the lateral molecular site within the row, resembling charging/discharging diamonds as a function of sample bias and a gate voltage (here the molecular site). The asymmetry of the molecular
junction in the STM configuration is reflected in the asymmetry of the
Coulomb diamonds.}
\end{figure}

The existence of two ground states with different electron occupation is caused
by subtle variations in the chemical equilibrium of TCNQ molecules with their
environment. Inhomogeneities of the surface potential induce a modulation of
the LUMO alignment with respect to the Fermi level \cite{note5} that eventually
can lead to a change of the charge state along the molecular rows. Fig. 4a
shows a dI/d\Vs(\Vs,x) map collecting a series of spectra taken along eight
TCNQ molecules embedded in an extended row. Accordingly, the spectral map shows
a gradual shift of their features with the lateral position along the chain.
The left most molecules are Type II (neutral) species (compare with individual
dI/d\Vs\ plots extracted in Fig. 4b). Their spectra show a charging dip
(marked by line $E^{-}$  in Fig. 4a) and the onset of the interface state (IS)
at negative bias, as well as the unoccupied LUMO state (E$_{LUMO}$ line)
slightly above \Ef. All these features gradually shift towards \Ef\ as we
explore molecules towards the right. A cross-over to a different ground state
occurs at the fifth molecule, where the three spectroscopic features cross
through \Ef. TCNQ molecules to the right of the fifth molecule lie in an
anionic ground state. The line  $E^{-}$   separating the neutral  and charged
state is now at positive bias, indicating the discharging peaks of the TCNQ
anions. At the same time the LUMO resonance vanishes as it becomes singly
occupied \cite{note9} and the interface state (IS line) becomes depopulated.
This indicates that the interface state provides the charge required for the
stabilization of the TCNQ anion.

Following the DBTJ model, the map in Fig. 4a allows us to obtain the bias
fraction $\alpha$  dropping at the molecule-metal interface. For Type II
molecules, $\alpha$ corresponds to the ratio $E_{LUMO}/(E^{-}-E_{LUMO})$,
amounting to $\sim 0.15$ for the spectra in Fig. 4 \cite{note6}. This value is
comparable with the one reported for molecules on alumina films
\cite{NazinPRL05}. The small vacuum barrier underneath the TCNQ molecules
behaves similar to thin insulating films with larger relative permeabilities

Below the $E^{-}$ line, a second conductance dip indicates the onset for doubly
charging the TCNQ molecules ($E^{2-}$ line). The bias separation between the
two charging lines ($\sim$ 1.2 V), corrected by the factor $\alpha$,
corresponds to a Coulomb charging energy U $\sim$ 180 meV. This value lies in
the expected range for molecules on metal surfaces and is much smaller than the
U in organic solids due to the charge screening effect of the underlying metal
surface \cite{TorrenteJPCM08}.

The gradual shifts of the onset for charging/discharging with the position of
the molecules within the chain resembles the transport characteristics of
three-terminal quantum devices (as sketched in Fig. 4c), where the alignment of
the quantum levels are tuned externally by a gate potential
\cite{Kouwenhoven01, JarilloNature04, BrarNatphys11}. In these devices, the
different charging states are represented as a set of Coulomb Diamonds
appearing in spectral maps as a function of the applied gate and source-drain
potential. In our experiments, the molecular charge state is shown to be tuned
in two different ways: i) the finite voltage drop at the molecule-metal
interface mimics a gating potential, which can be tuned by the tip-sample
distance. ii) the local adsorption potential energy surface acts as a "static"
gate potential, tuning various charge states depending on the lateral position
along molecular TCNQ rows (as in Fig. 4).

In summary, our results show that charge localization in molecular layers
directly on a metal is possible at the level of reaching integer molecular
charge states, which can be manipulated. We observed that TCNQ molecules may be
found with two distinct oxidation states (anionic and neutral),  depending on
small variations of the underlying adsorption potential. These charge states
are crucially affected by externally applied local electric fields, allowing
tuning the molecular tunneling conductance by field effect.

We thank  Jordi Fraxedas for driving us into this study, and Felix von Oppen and Tobias Umbach for stimulating discussions. This
research has been supported by the DFG through grants FR2726, SPP1243 and
Sfb658.


\begin{thebibliography}{99}
\bibitem{note1} The Au(111) surface was cleaned by repetitive cycles of
    sputtering and annealing. Tungsten tips were prepared in situ by gently
    dipping them into the gold substrate until they show no tip-related
    spectroscopic feature. Sublimation temperatures of TMTTF and TCNQ are
    90$^{\circ}$C and 140$^{\circ}$C, respectively.
\bibitem{note7} We note that  these effects were not observed when using
    instead the parent specie TTF, but only after the incorporation of  methyl
    terminations in the TTF backbone. This suggests that the methylation
    induces a larger separation of the TCNQ plane from the surface and,
    consequently, enhances the DBTJ behaviour.
\bibitem{note3} It is remarkable that, upon charging Type II
    molecules at negative bias, the blockade induced by the formation
    of a local dipole is more important than the expected increase in
    tunneling transmission due to an additional resonance (the now
    singly occupied orbital) below \Ef.
\bibitem{note5}   These variations are probably originated by the stress caused
    by the underlying Au(111) herring-bone reconstruction. Regions with many
    dislocation lines show only the anionic species and no reconstruction.
\bibitem{note9} The two single particle peaks at SOMO and SOMO+U are expected
    to have less intensity in the spectra than the corresponding fully
    unoccupied (LUMO) resonance and, then, difficult to resolve \cite{Temirov}.
\end{thebibliography}
 \end{document}